\documentclass[reprint,amsmath,amssymb,aps]{revtex4-1}

\usepackage{color}
\usepackage{amsmath}
\usepackage{graphicx}
\usepackage{dcolumn}
\usepackage{bm}
\usepackage{booktabs}
\usepackage{multirow}

\begin{document}
\preprint{APS/123-QED}

\title{Underwater Transmission of High-dimensional Twisted Photons over 55 Meters}

\author{Yuan~Chen$^{1,2}$, Wei-Guan~Shen$^{1,2}$, Zhan-Ming~Li$^{1,2}$,  Cheng-Qiu~Hu$^{1,2}$, Zeng-Quan~Yan$^{1,2}$, Zhi-Qiang~Jiao$^{1,2}$, Jun~Gao$^{1,2}$, Ming-Ming~Cao$^{1}$, Ke~Sun$^{1}$}
\author{Xian-Min Jin$^{1,2,}$}
\email{xianmin.jin@sjtu.edu.cn}

\affiliation{
	$^1$School of Physics and Astronomy, Shanghai Jiao Tong University, Shanghai 200240, China\\
	$^2$Synergetic Innovation Center of Quantum Information and Quantum Physics, University of Science and Technology of China, Hefei, Anhui 230026, China}
\date{\today}

\maketitle

\textbf{As an emerging channel resource for modern optics, big data, internet traffic and quantum technologies, twisted photons carrying orbital angular momentum (OAM) have been extended their applicable boundary in different media, such as optical fiber and atmosphere. Due to the extreme condition of loss and pressure, underwater transmission of twisted photons has not been well investigated yet. Especially, single-photon tests were all limited at a level of a few meters \cite{Ji2017,Bouchard2018}, and it is in practice unclear what will happen for longer transmission distances. Here we experimentally demonstrate the transmission of single-photon twisted light over an underwater channel up to 55 meters, which reach a distance allowing potential real applications. For different order OAM states and their superposition, a good preservation of modal structure and topological charge are observed. Our results for the first time reveal the real transmission performance of twisted photons in a long-distance regime, representing a step further towards OAM-based underwater quantum communication.}

\section*{INTRODUCTION}
In virtue of the unique characteristics of twisted light \cite{Allen1992}, the application boundary has been extended from remote sensing \cite{Fickler2012}, light detection \cite{Lavery2013}, classical optical communication \cite{Wang2012,Bozinovic2013,Baghdady2016,Ren2016,Zhao2017}, to high-dimensional quantum information processing \cite{Dada2011,Mirhosseini2015,Bouchard2017,Erhard2018}. As an additional degree of freedom and an emerging resource for both classical and quantum information technologies, the propagation of twisted light is regarded as a crucial problem to be solved \cite{Zhao2015}. Many experiments have been carried out  to obtain the dynamics and propagation properties of twisted light in different media, such as in fiber \cite{Bozinovic2013,Cozzolino2018} and atmosphere \cite{Krenn2014,Krenn2016,Lavery2017}. 

During propagation in free-space open air, the wavefront of light gets distorted due to atmospheric turbulence \cite{Rodenburg2012}. The variation of atmospheric temperature and pressure with time results in the random change of refractive index, which further leads to the phase distortion of the transmitted beam. In contrast, free-space water is a uniform isotropic medium, though incorporating dissolved salts and microbes, does not lead to massive polarization rotation or depolarization of single photons\cite{Ji2017}. The isotropic water also induces very limited refractive index change \cite{Austin1976}. These imply that free-space water potentially can be a promising channel for delivering both photonic polarization and OAM quantum states, especially in the relatively quiet environment of deep sea. 

Comparing with optical fiber and atmosphere, underwater channel has a much more extreme physical environment, which makes it challenging to access experimentally. Beside high loss, high pressure and slow classical communication rate, one of immediate difficulties, technically for example, is to steer the photon stream in water and precisely point to the receiving terminal. Thus, while very high channel capacities of data rate have been demonstrated for classical underwater optical communication, the achieved distance are still limited \cite{Baghdady2016,Ren2016}.

Quantum communication employs single photons to encode information, therefore is more sensitive to all type of underwater harmful effects than classical communication. Ji et al presents the first quantum communication experiment through seawater, which confirms that the polarization degree of freedom of single photons can be preserved well when propagated through 3.3-meter-long seawater channel \cite{Ji2017}. Very recently, Bouchard et al performed an OAM-based quantum key distribution (QKD) through 3-meter underwater channel and explored the effect of turbulence on transmitted error rates \cite{Bouchard2018}. 

However, it is still unclear what will happen if single photons go through longer transmission distances, and what tolerance the degree of freedom can have. The theoretically achievable communication distances obtained by different single-photon scattering models are even not consistent with one another \cite{Ji2017,Shi2014}. It is therefore very demanding to push the single-photon test towards an entirely new region, for not only fundamentally verifying theoretical models but also boosting immediate applications in special and ultimate scenarios.

\begin{figure*}
\centering
\includegraphics[width=2\columnwidth]{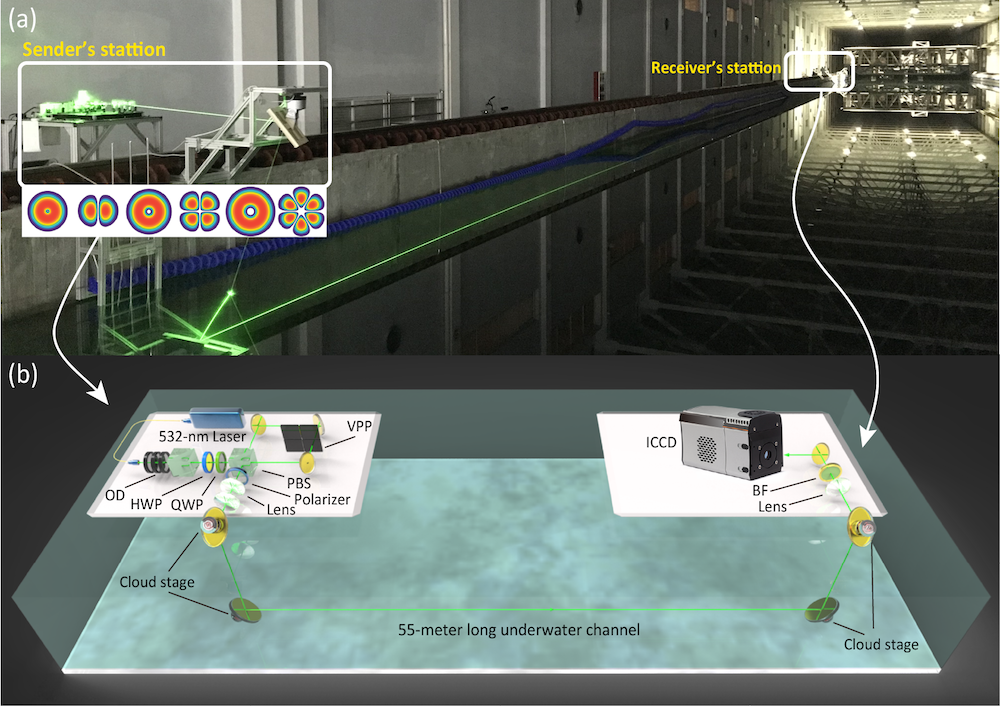}
\caption{\textbf{Experimental implementation.} \textbf{a.} Real field-test environment. \textbf{b.} Sketch of the experimental setup. The single-photon twisted light is generated using a modified Sagnac loop with a vortex phase plate (VPP) placed inside the loop. The Sagnac loop can operate in both single-directional and bidirectional circulation by preparing the polarization of incident photons with a half wave plate (HWP) and a quarter wave plate (QWP) before the loop. After superposition states or pure states going through the underwater channel, their modal structures are measured with an ICCD camera. The inset shows typical classes of OAM states adopted for experimental tests. OD: neutral-density filter; PBS: polarization beam splitter; f$_{1}$-f$_{3}$: lens; BF: band-pass filter (532nm $\pm$ 10nm); ICCD: intensified charge coupled device camera.}
\label{Figure 1}
\end{figure*}

\begin{figure*}
\centering
\includegraphics[width=2\columnwidth]{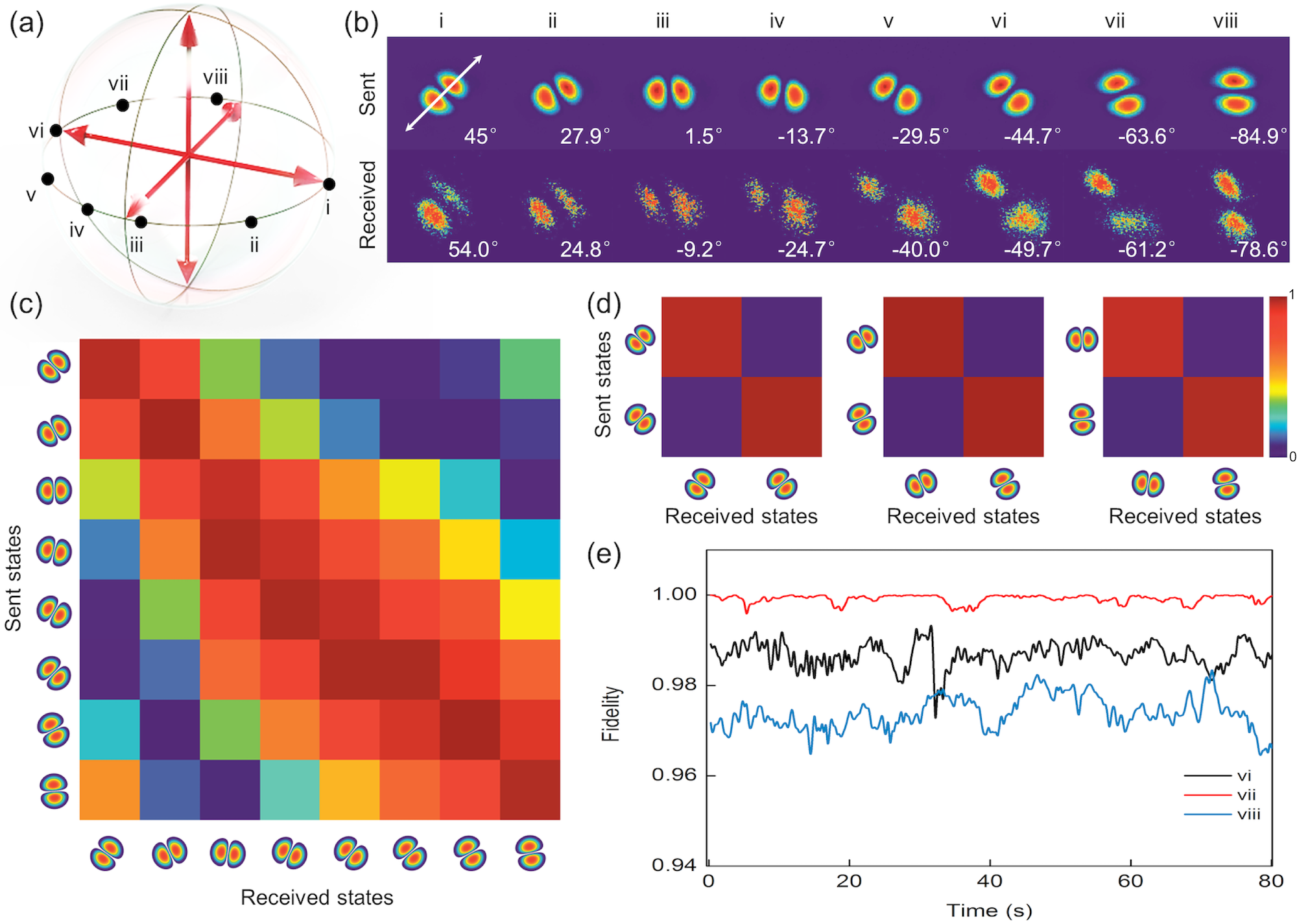}
\caption{\textbf{Experimental results of first-order superposition states.} \textbf{a.} Superposition states presented on Bloch sphere. \textbf{b.} Measured modal structures of sent and received states with their identified relative phases. The double-headed arrow illustrates the identified orientation of modal structure with our angle recognition algorithm. It should be noticed that the identified orientation of modal structure reflects the relative phase of superposition state but does not equal to $\theta$. \textbf{c.} Cross-talk matrix obtained by projecting each received state on all the sent states. The sent and received states are labeled by simulated modal structures for straightforward viewing. \textbf{d.} Projections on the mutually orthogonal states. \textbf{e.} Real-time fidelity measurement of first-order superposition states vi, vii, viii.}
\label{Figure 2}
\end{figure*}

\begin{figure}
\centering
\includegraphics[width=1\columnwidth]{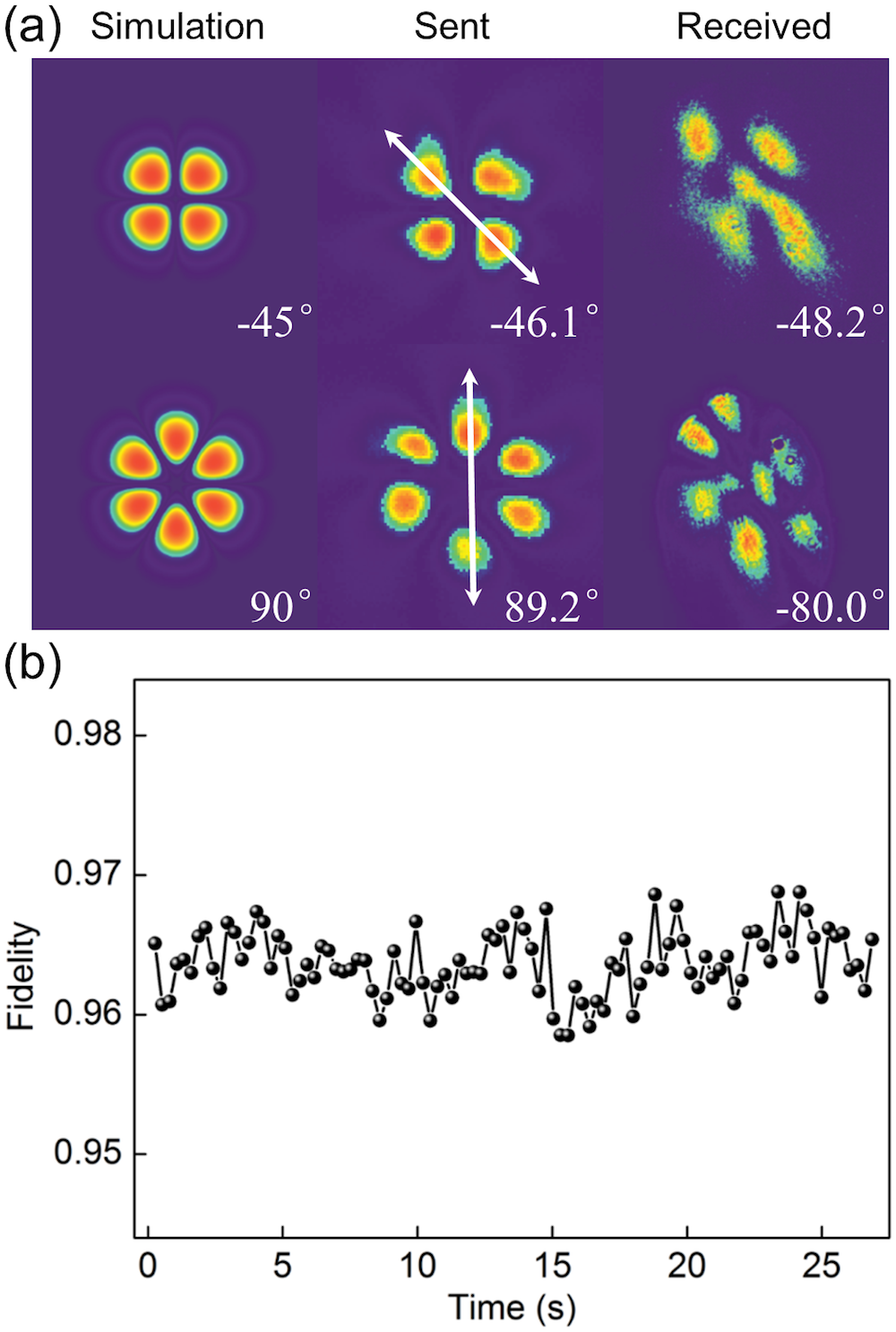}
\caption{\textbf{Experimental results of higher-order superposition states}. \textbf{a.} Simulation and measured modal structures of second- and  third-order superposition states. The double-headed arrow illustrates the identified orientation of modal structure with our angle recognition algorithm. \textbf{b.} Real-time fidelity measurement of third-order superposition states. The fluctuation of fidelities over 27 seconds is found less than 1.1\%.}
\label{Figure3}
\end{figure}

\begin{figure}
\centering
\includegraphics[width=1\columnwidth]{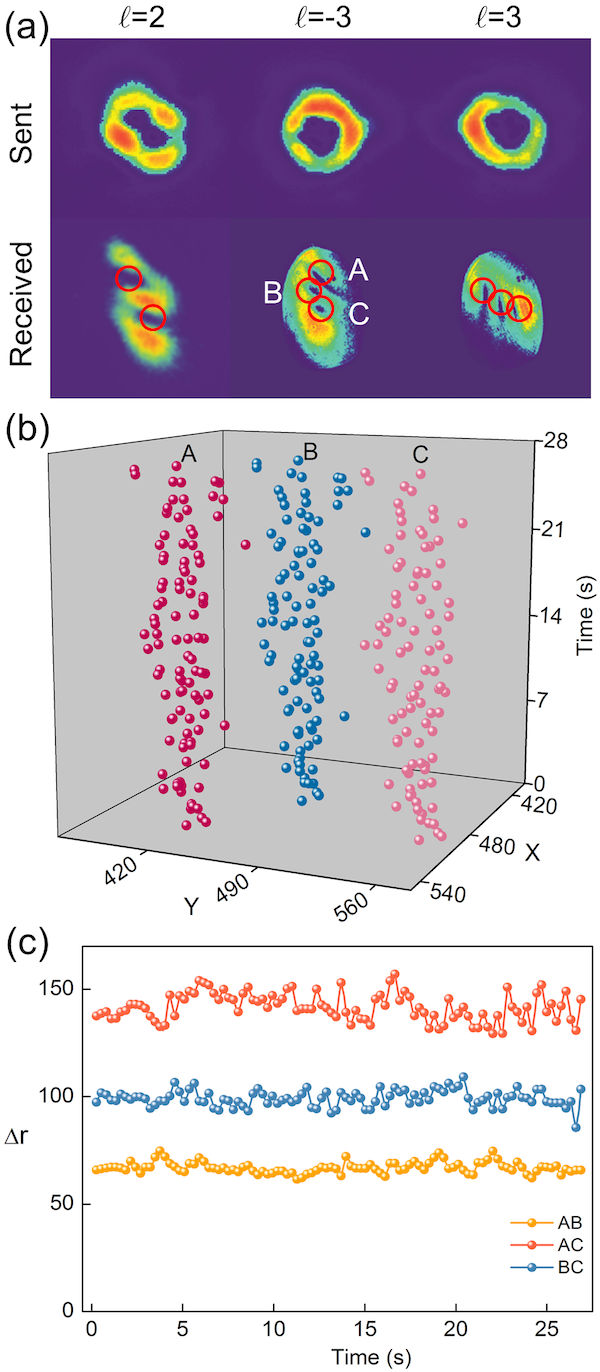}
\caption{\textbf{Propagation dynamics of high-order vortices}. \textbf{a.} Measured modal structures of high-order pure states. The red circles mark the unit-strength vortices split from a high-order vortex core. \textbf{b.} The evolution of the positions of three vortex cores over time for $\ell=-3$ pure state. The label X and Y axis represent the absolute position, so there are no units. \textbf{c.} The evolution of the inter-core distance ($\Delta r$: pixel in ICCD) for $\ell=-3$ pure state.}
 \label{Figure4}
\end{figure}

In this letter, we experimentally demonstrate the transmission of single-photon twisted light over a underwater channel up to 55 meters, about 19 times longer than previous record \cite{Ji2017,Bouchard2018}. By an angle recognition algorithm, we find that the modal structure for different order OAM superposition states can be maintained well with a fidelity over 96\%. In the presence of perturbation, a high-order vortex will break up into individual vortex. In our experiment, we also observe this phenomenon and find that the number of topological charge can be kept. The maintenance of modal structure and the number of topological charge to some extent indicates that the underwater channel, especially in a relatively quiet environment, may be promising for twisted-photon-encoded quantum communication. 

\section*{RESULTS}
A schematic layout of the experimental setup is shown in Fig. 1. We prepare single photons by strongly attenuating the solid-state 532-nm laser to about 0.51 photon per time slot (1ns) before entering water interface, which is a typical parameter adopted in decoy state protocols \cite{Lo2005,Peng2007}. We encode the high-dimensional OAM degree of freedom into photons by using a modified Sagnac loop equipped with a vortex phase plate (VPP). 

The Sagnac loop can operate in both single-directional and bidirectional circulation by preparing the polarization of incident photons with the two wave plates shown in Fig. 1. Single photons initialized in horizontal ($H$) and vertical ($V$) polarization run in single-directional circulation, generating pure state $\arrowvert{-\ell}\rangle$ and $\arrowvert{\ell}\rangle$ respectively with a corresponding VPP inside the loop. Single photons initialized in superposition $1/\sqrt{2}(\arrowvert{H}\rangle+e^{i\theta}\arrowvert{V}\rangle)$ polarization run in bidirectional circulation and generate a hybrid entangled state $1/\sqrt{2}(\arrowvert{H{-\ell}}\rangle+e^{i\theta}\arrowvert{V{\ell}}\rangle)$. After projection on diagonal polarization, we finally can obtain superposition state ${1}/{\sqrt2}(\arrowvert{-\ell}\rangle+e^{i\theta}\arrowvert{\ell}\rangle)$. Fig. 2a gives one class of the prepared superposition states for $\ell=1$ in a Bloch sphere \cite{Jack2010}, equivalent to the Poincar\'e sphere for polarization \cite{Padgett1999}. 

The superposition states shown in the Bloch sphere with different relative phases $\theta$ result in different orientations of the modal structures, which means they can be distinguished according to their direction. The twisted photons pass through 55-meter-long water channel located underneath 1.5 meter of the surface of water. By selecting lenses for telescope systems in both sides, we are able to obtain an appropriate spot size at the receiving terminal. To characterize the received mode, we measure the modal structures of twisted photons by using a single-photon sensitive ICCD camera. We develop an angle recognition algorithm to analyze the obtained images and retrieve the information encoded in the relative phases. To be specific, the angle recognition algorithm firstly precisely identify the the centroid of each petal, and then link the centroids to determine the angle according to the x-axis.

The measured modal structures and the retrieved relative phases are shown in Fig. 2b. The deviations are found restricted between $2.4^\circ$ and $11.0^\circ$. The average deviation is $7.2^\circ$ and the relatively small deviation can be used to benchmark the ability of preserving the spatial structure of OAM. By projecting each received state onto all the sent state bases, we can calculate the cross-talk matrix (see Fig. 2c). Apparently, the crosstalk naturally depends on how overlapped the two state bases are, no matter whether channel errors exist. We can see that, while there some deviations measured, this basic trend of the crosstalk are found very clear and reasonable. A small crosstalk can be obtained when the difference between the two state bases goes larger than $22.0^\circ$, which is consistent with previous results obtained in free-space air \cite{Krenn2014}.

The diagonal of the cross-talk matrix gives the projection results of the received states to their corresponding sent states. The measured fidelities for the prepared superposition state are all over 96\%. In our experiment, six of the eight sent states, except iv and v, suppose to orthogonal to each other pair by pair, though there exists some imperfection for preparing the sent states. We present these pair-like projections of the mutually orthogonal states in Fig. 2d. The strong contrast well visibly indicates a good preservation of encoded relative phases via the twisted photons through long-distance underwater channel.

In order to verify the stability of the underwater channel, we send three higher photon-flux classical light in superposition states vi, vii, viii. By continuously collecting and analyzing real-time images, we can obtain the changes of the fidelities over time. Counterintuitively, the fluctuations of the measured fidelities are very small, only in a range of 0.5\% to 2.8\%, at least it is true in a relatively quiet environment.

We then experimentally investigate the transmission properties of higher-order OAM states. While we also perform the single-photon test for the second-order twisted photons, we prepare the third-order twisted light in a higher photon flux in order to probe both the fluctuation of fidelity and the dynamics of vortex core in real time. The simulation and measured modal structures of the second-order superposition state are shown in Fig. 3a. With the angle recognition algorithm, the obtained fidelities for both second- and third-order states are larger than 96\%. The fluctuation of fidelities over 27 seconds is found less than 1.1\%. These results once again verify that the higher-order superposition states can be maintained after passing through long-distance underwater channel.

We further look into the propagation dynamics of vortex cores for pure state, which may be highly relevant to the fields of classical and quantum communication. Ideally, paraxial beam with a well-defined topological charge has a cylindrically symmetric intensity profile. However, breaking the cylindrically symmetry makes the higher-order phase singularities tend to split into sets of lower-order singularities \cite{Mamaev1997,Dennis2006,Kumar2011,Bahl2014}. The turbulence of water can induce random refractive index change on the cross section of channel as well as along the propagation direction, and therefore naturally break the cylindrically symmetry. As is shown in Fig. 4a, we do observe that the higher-order vortex core splits into sets of unit-strength vortices.

Interestingly, we find that, in spite of the turbulence induced vortex splitting, there always exists three vortices core for the third-order OAM state shown in Fig. 4b. While we can clearly see the beam wandering from the evolution of the position of each vortex core over time, we find that their inter-core distances are considerably stable (see Fig. 4c). In addition, our results reveal that, although the higher-order vortex core breaks up into sets of unit-strength vortices, the number of vortex cores equals to the order of twisted photons and preserves well. The observed conservation and robustness provide a potential approach to transmit information through long turbulence underwater channel.

\section*{DISCUSSION}
In summary, we present the first experimental attempt to explore the transmission property of the single-photon twisted light over a long underwater channel. Radio waves can only penetrate meters of water, which makes underwater area a forbidden region for the radio signal from satellites or aircrafts. Our experimental demonstration of transmission of twisted photons shows the feasibility of quantum communication with submersibles up to 55 meters, and may promise even longer achievable distances with high photon-flux sent states in classical regime.  

In addition, it is well known that the density of seawater associated with the refractive index varies with temperature, salinity and pressure. In deep sea, these parameters remain essentially constant, and as a consequence the density too does not change much, which indicates that the refractive index changes very little \cite{Austin1976}. This implies that some ultimate environments like deep sea might be a good channel for delivering twisted photons in a finite distance. Our test in relatively quiet underwater over 55 meters represents a step further towards submarine quantum communication and quantum sensing with twisted photons.

\section*{Methods}
\subsection{\textbf{Experimental details:}}
The whole experiment is conducted in a marine test platform located in the University of Shanghai Jiao Tong University. The marine test platform is the biggest multiple function towing tank in Asia, and is semi-open with many windows connecting the internal and external environment. We install the sender and the receiver systems on the same side of water. We guide photons from the sender system into water by a wireless-controlled cradle head, and collect photons back into free-space air with the receiver system. The optical link consists of two air-water interfaces and a long underwater channel up to 55 meters. We use a 532 nm laser with a tunable power range of 0-100 mW as the light source because a blue-green band of 400 nm-550 nm light suffers less attenuation in water \cite{Wozniak2007}. We prepare the initial states at single-photon level by setting the output power as low as $1.898\times10^{-19}$ W, which is equivalent to a source with an average photon number of 0.51 per nanosecond.  We employ a high power beacon light with a power up to 10 W to align and optimize the underwater channel. Alignment at 1.5-meter-deep underwater environment is challenging as we need the precise pointing control with cradle head and mirrors to the receiving terminal. 

\begin{figure}
\centering
\includegraphics[width=1\columnwidth]{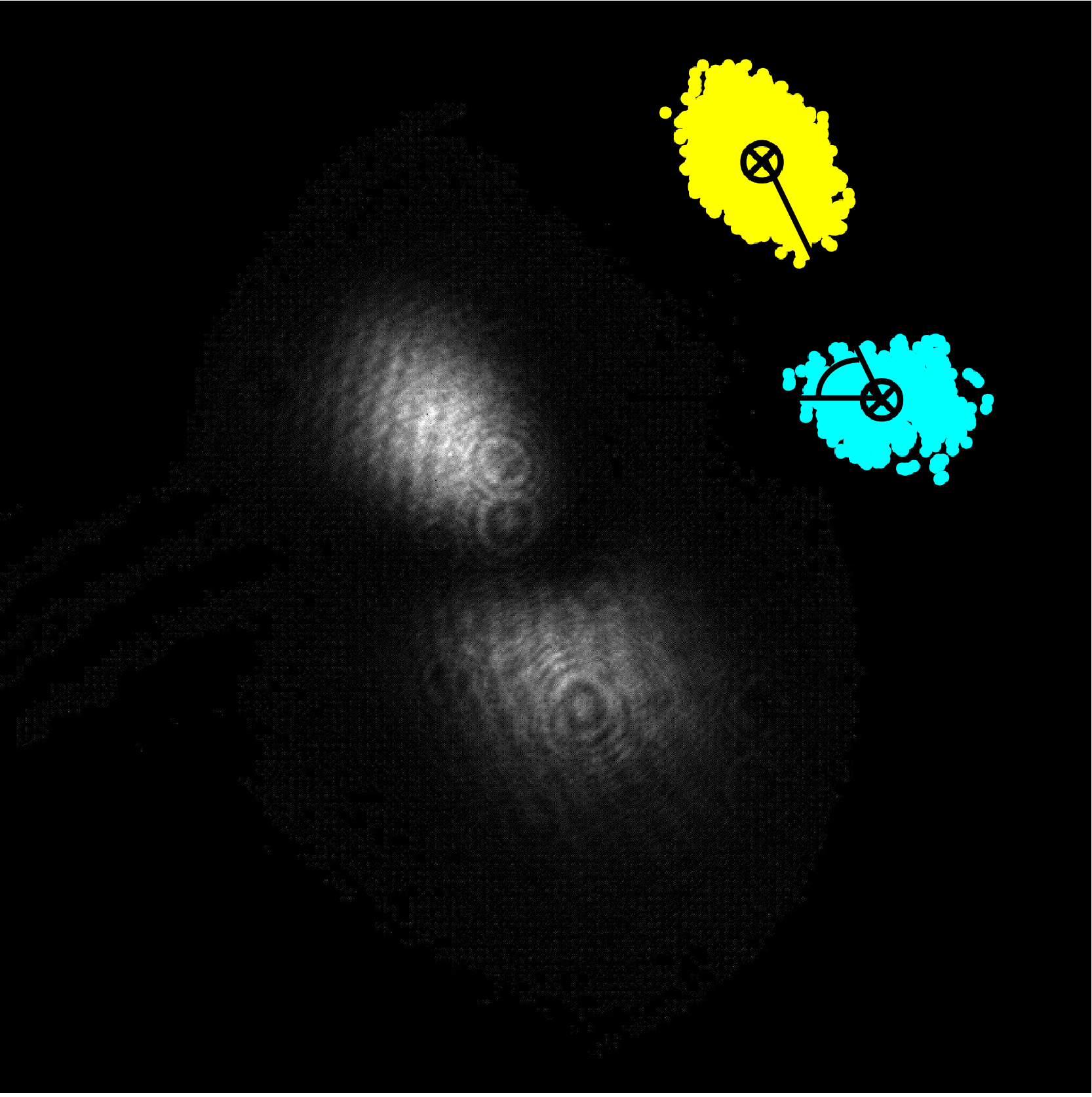}
\caption{\textbf{An example of the calculation of the angle}. The gray-scale image is the original first-order superposition state, and the inset shows the centroid of two petals by the angle recognition algorithm, which is used to calculate the fidelity.}
\label{Figure 5}
\end{figure}

\begin{figure}
\centering
\includegraphics[width=1\columnwidth]{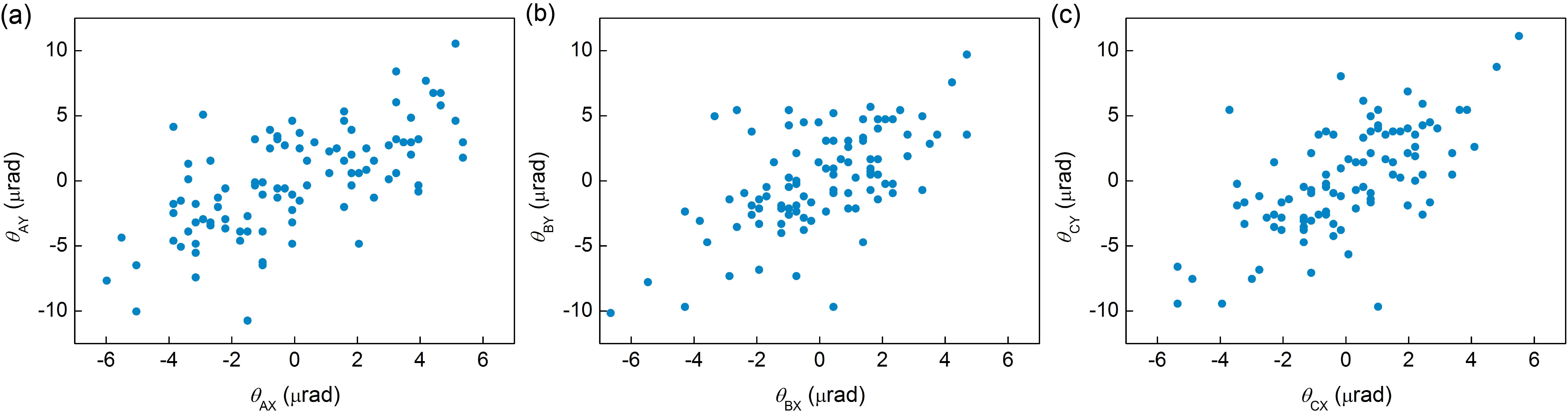}
\caption{\textbf{Tip-tilt effect in underwater channel}. \textbf{a.} Core A variation from an on-axis position measured over 20 s. \textbf{b.} Core B variation from an on-axis position measured over 20 s. \textbf{c.} Core C variation from an on-axis position measured over 20 s.}
\label{Figure 6}
\end{figure}

By measuring the power at both the sending terminal and receiving terminal, we are able to estimate the channel loss. The measured attenuation coefficient is about 0.16 m$^{-1}$, which is close to the coefficient in coastal seawater (0.179 m$^{-1}$) \cite{Cochenour2008}. The loss of the overall system is about 40 dB. In spite of high loss, we still can retrieve the weak signal from very strong background noises via a single-photon sensitive ICCD. In addition, we employ the Sagnac-like interferometric setup instead of a spatial light modulator (SLM) to prepare all the initial modes in the experiment. Not only because this solution is more economical, but also the configuration potentially provide a way of fast state encoding. We can realize fast modulation of the initial states if we replace the wave plate outside of the interferometer with a Pockels cell. 

\subsection{\textbf{Fidelity estimation:}} The superposition states shown in the Bloch sphere with different relative phases result in different orientations of the modal structures, which means they can be distinguished according to their direction. We employ the angle recognition algorithm to analyze the obtained images. Firstly, we precisely identify the the centroid of each petal, and then link the centroids to determine the angle according to the x-axis (see Fig. 5). It should be noticed that the superposition states prepared on the Bloch sphere with different relative phases is equivalent to the Poincar\'e sphere for polarization \cite{Padgett1999}. Therefore, we can project each received state onto the corresponding sent state base to calculate the fidelity. Apparently, the fidelity naturally depends on how overlapped the two state bases are.

\subsection{\textbf{Estimation of water turbulence:}}The tip-tilt aberration arising from water turbulence is one of the concerns within a free-space water optical system \cite{Bouchard2018}. The aberration results in a change of the beam propagation direction. One can characterize the degree of tip-tilt by measuring the centroid location of a received Gaussian beam \cite{Lavery2017}. Similarly, here we can measure the variation from an on-axis position of each vortex core over 20 s by the ICCD, whose exposure time is set 0.03 s. The position variation across the beam axis is angled at ${\rm \theta_{X}=\arctan(\Delta{x}/L)}$, ${\rm \theta_{Y}=\arctan(\Delta{y}/L)}$ for tip and tilt, respectively, where L is the length of the underwater link. We observe an average radial variation of 0.64 mm from the center (see Fig. 6). 

\subsection*{Acknowledgments}
The authors thank Jian-Wei Pan for helpful discussions. This work was supported by the National Key R\&D Program of China (Grant No. 2017YFA0303700), the National Natural Science Foundation of China (Grants No. 61734005, No. 11761141014, and No. 11690033), the Science and Technology Commission of Shanghai Municipality (Grants No. 15QA1402200, No. 16JC1400405, and No. 17JC1400403), the Shanghai Municipal Education Commission (Grants No. 16SG09 and No. 2017-01-07-00-02-E00049), and the Zhiyuan Scholar Program (Grants No. ZIRC2016-01). X.-M.J. acknowledges support from the National Young 1000 Talents Plan.


\begin{thebibliography}{99} 
{
\bibitem{Ji2017}Ji, L. {\it et al.} Towards quantum communications in free-space seawater. {\it Opt. Express} {\bf 25}, 19795-19806 (2017).
\bibitem{Bouchard2018}Bouchard, F. {\it et al.} Quantum cryptography with twisted photons
through an outdoor underwater channel. {\it Opt. Express} {\bf 26}, 22563-22573 (2018).
\bibitem{Allen1992}Allen, L. {\it et al.} Orbital angular momentum of light and the transformation of Laguerre-Gaussian laser modes. {\it Phys. Rev. A} {\bf 45}, 8185 (1992).
\bibitem{Fickler2012}Fickler, R. {\it et al.} Quantum entanglement of high angular momenta. {\it Science} {\bf 338}, 640-643 (2012).
\bibitem{Lavery2013}Lavery, M. P. J. {\it et al.} Detection of a spinning object using light's orbital angular momentum. {\it Science} {\bf 341}, 537-540 (2013).
\bibitem{Wang2012}Wang, J. {\it et al.} Terabit free-space data transmission employing orbital angular momentum multiplexing. {\it Nat. Photon.} {\bf 6}, 488-496 (2012).
\bibitem{Bozinovic2013}Bozinovic, N. {\it et al.} Terabit-scale orbital angular momentum mode division multiplexing in fibers. {\it Science} {\bf 340}, 1545-1548 (2013).
\bibitem{Baghdady2016}Baghdady, J. {\it et al.} Multi-gigabit/s underwater optical communication link using orbital angular momentum multiplexing. {\it Opt. Express} {\bf 24}, 9794-9805 (2016).
\bibitem{Ren2016}Ren, Y.-X. {\it et al.} Orbital angular momentum-based space division multiplexing for high-capacity underwater optical communications. {\it Sci. Rep.} {\bf 6}, 33306 (2016).
\bibitem{Zhao2017}Zhao, Y.-F. {\it et al.} Performance evaluation of underwater optical communications using spatial modes subjected to bubbles and obstructions. {\it Opt. Lett.} {\bf 42}, 4699-4702 (2017).
\bibitem{Dada2011}Dada, A. C. {\it et al.} Experimental high dimensional two-photon entanglement and violations of generalized Bell inequalities. {\it Nat. Phys.} {\bf 7}, 677-680 (2011).
\bibitem{Mirhosseini2015}Mirhosseini, M. {\it et al.} High-dimensional quantum cryptography with twisted light. {\it New J. Phys.} {\bf 17}, 033033 (2015).
\bibitem{Bouchard2017}Bouchard, F. {\it et al.} High-dimensional quantum cloning and applications to quantum hacking. {\it Sci. Adv.} {\bf 3}, e1601915 (2017).
\bibitem{Erhard2018}Erhard, M. {\it et al.} Twisted photons: new quantum perspectives in high dimensions. {\it Light: Sci. Appl.} {\bf 7}, 17146 (2018). 
\bibitem{Zhao2015}Zhao, N.-B. {\it et al.} Capacity limits of spatially multiplexed free-space communication. {\it Nat. Photon.} {\bf 9}, 822-828 (2015).
\bibitem{Cozzolino2018}Cozzolino, D. {\it et al.} Fiber based high-dimensional quantum communication with twisted photons. arXiv preprint arXiv:1803.10138 (2018).
\bibitem{Krenn2014}Krenn, M. {\it et al.} Communication with spatially modulated light through turbulent air across Vienna. {\it New J. Phys.} {\bf 16}, 113028 (2014).
\bibitem{Krenn2016}Krenn, M. {\it et al.} Twisted light transmission over 143 km. {\it Proc. Natl. Acad. Sci.} {\bf 113}, 13648-13653 (2016).
\bibitem{Lavery2017}Lavery, M. P. J. {\it et al.} Free-space propagation of high-dimensional structured optical fields in an urban environment. {\it Sci. Adv.} {\bf 3}, e1700552 (2017).
\bibitem{Rodenburg2012}Rodenburg, B. {\it et al.} Influence of atmospheric turbulence on states of light carrying orbital angular momentum. {\it Opt. Lett.} {\bf 37}, 3735-3737 (2012).
\bibitem{Austin1976}Austin, R. W., Halikas, G. The index of refraction of seawater. SIO/University of California, SIO No. 76-1 (1976).
\bibitem{Shi2014} Shi, P., Zhao, S. C., Li, W. D. \& Gu, Y. J. Preprint at http://arxiv.org/abs/1402.4666v2 (2014).
\bibitem{Lo2005}Lo, H. K. {\it et al.} Decoy state quantum key distribution. {\it Phys. Rev. Lett.} {\bf 94}, 230504 (2005).
\bibitem{Peng2007}Peng, C. Z. {\it et al.} Experimental long-distance decoy-state quantum key distribution based on polarization encoding. {\it Phys. Rev. Lett.} {\bf 98}, 010505 (2007).
\bibitem{Jack2010}Jack, B. {\it et al.} Entanglement of arbitrary superpositions of modes within two-dimensional orbital angular momentum state spaces. {\it Phys. Rev. A} {\bf 81}, 043844 (2010).
\bibitem{Padgett1999}Padgett, M. J. {\it et al.} Poincar\'e-sphere equivalent for light beams containing orbital angular momentum. {\it Opt. Lett.} {\bf 24}, 430-432 (1999).
\bibitem{Mamaev1997}Mamaev, A. V. {\it et al.} Decay of high order optical vortices in anisotropic nonlinear optical media. {\it Phys. Rev. Lett.} {\bf 78}, 2108 (1997).
\bibitem{Dennis2006}Dennis, M. R. Rows of optical vortices from elliptically perturbing a high-order beam. {\it Opt. Lett.} {\bf 31}, 1325-1327 (2006).
\bibitem{Kumar2011}Kumar, A. {\it et al.} Crafting the core asymmetry to lift the degeneracy of optical vortices. {\it Opt. Express} {\bf 19}, 6182-6190 (2011).
\bibitem{Bahl2014}Bahl, M., Senthilkumaran, P. Focal plane internal energy flows of singular beams in astigmatically aberrated low numerical aperture systems. {\it J. Opt. Soc. Am. A} {\bf 31}, 2046-2054 (2014).
\bibitem{Wozniak2007}Wozniak, B., Dera, J. Light absorption in sea water, vol. 33 (Springer, 2007).
\bibitem{Cochenour2008}Cochenour, B. M., Mullen, L. J., Laux, A. E. Characterization of the beam-spread function for underwater wireless optical communications links. {\it IEEE Journal of Oceanic Engineering} {\bf 33}, 513-521 (2008).
}
\end{thebibliography}
\end{document}